\documentclass{article}
\title{Reconstruction of higher stage first-class constraints into the secondary ones}
\author{A. A. Deriglazov\footnote{alexei.deriglazov@ufjf.edu.br ~ On leave of
absence from Dept. Math. Phys., Tomsk Polytechnical University,
Tomsk, Russia.}}
\date{Dept. de Matematica, ICE, Universidade Federal de Juiz de Fora,\\
MG, Brazil; \\
and \\
LAFEX - CBPF/MCT, Rio de Janeiro, RJ, Brazil.}
\begin{document}
\maketitle
\large

\begin{abstract}
We analyze a singular theory with first class constraints of an arbitrary stage.
Relation among the formulations of the constrained
system in terms of complete and extended Hamiltonians is clarified. We replace the extended Hamiltonian $H_{ext}$ by an improved one. The improved Hamiltonian has the same structure as $H_{ext}$ (higher stage constraints enter into the Hamiltonian in the manifest form), but, in contrast to $H_{ext}$, it arises as the complete Hamiltonian for some Lagrangian $\tilde L$, called the extended Lagrangian. This implies, in particular, that all the quantities appearing in the improved Hamiltonian have a clear meaning in the Dirac framework. $\tilde L$ is obtained in a closed form in terms of quantities of the initial formulation $L$. The formulations with $L$ and $\tilde L$ turn out to be equivalent. As an application of the formalism, we found local symmetries of $\tilde L$ in a closed form. All the constraints of $L$ turn out to be gauge symmetry generators for $\tilde L$. The procedure is illustrated with an example of a model with fourth-stage constraints.
\end{abstract}

\noindent

\section{Introduction}
The singular Lagrangian theory generally has a complicated structure of Lagrangian equations, which may consist of both second and first-order differential equations as well as algebraic ones. Besides, when some local symmetries are present, there may be identities among the equations that imply functional arbitrariness in the corresponding solutions. It should be mentioned that, in the modern formulation, popular field theories (electrodynamics, gauge field theories, the standard model, string theory, etc.) are of this type. An analysis of the singular theory is usually carried out according to the Dirac procedure [1, 2] in a
Hamiltonian formalism. This gives a somewhat clearer geometric picture of the classical dynamics:
all the solutions are restricted to lie on some surface in the phase space (determined by Dirac constraints), while the above mentioned arbitrariness is removed by postulating classes of equivalent trajectories. Physical quantities are then represented by functions defined on the classes.

The basic object of the Hamiltonian formulation turns out to be the complete Hamiltonian $H$ $=$ $H_0$ $+$ $v^\alpha\Phi_\alpha$ that is uniquely constructed from the initial Lagrangian $L$ (see the next section for details). Here $H_0$ is the Hamiltonian, $v^\alpha$ represents primarily inexpressible velocities [2], and $\Phi_\alpha$ are primary constraints.
The application of the Dirac procedure can reveal higher-stage constraints denoted $T_a$. In this work the constraints are assumed to be first class, but they can be of arbitrary order: the constraints of order $N$ arise after imposing the time preservation of constraints of order $N-1$.

A further analysis of the constrained system can be carried out in the so called formalism of the extended Hamiltonian. The latter is obtained from the complete Hamiltonian by addition of all the higher-stage constraints with corresponding multipliers: $H_{ext}$ $\equiv$ $H$ $+$ $\lambda^aT_a$. Hamiltonian equations, corresponding to $H_{ext}$, involve derivatives of $T_a$ and hence are different from the equations obtained from $H$. Nevertheless, it can be proved that the two systems are equivalent.  The equivalence can be easily demonstrated for the case of second-class constraints (see section 2.3 in [2]), whereas the case of first-class constraints requires a rather detailed analysis in terms of canonical variables (section 2.6 in [2]).

Due to a special structure of $H_{ext}$ (all the constraints enter into $H_{ext}$ in the manifest form), the extended formulation proves to be a very useful tool for the analysis of both the general structure [2] and local symmetries [3] of the singular theory. At the same time, the origin of the extended Hamiltonian and its relation to the complete one remain somewhat mysterious. In particular, $H_{ext}$ cannot be treated as the complete Hamiltonian of some Lagrangian theory (see Sect. 3 for details), so the multipliers $v^a$ have no proper interpretation.

In the first part of the present work, we clarify the relation among the formulations of a constrained
system in terms of the complete and the extended Hamiltonians. Our suggestion is to improve the extended formalism by replacing $H_{ext}$ with an improved extended Hamiltonian $\tilde H$ (see Eq. (\ref{9}) below). It has the same structure as $H_{ext}$ (all the constraints enter into $\tilde H$ in the manifest form), but, in contrast to $H_{ext}$, it arises as the complete Hamiltonian for some Lagrangian $\tilde L$, called the extended Lagrangian. This implies, in particular, that all the quantities appearing in $\tilde H$ have a clear meaning in the Dirac framework. The extended Lagrangian $\tilde L$ is obtained in a closed form in terms of quantities of the initial formulation (see Eq. (\ref{15}) below). The formulations with $L$ and $\tilde L$ turn out to be equivalent, which implies a clear relation among the initial and the improved formulations in the Lagrangian framework. Due to the equivalence, it is a matter of convenience which formulation to use for description of the theory under
consideration\footnote{Let us also point out that higher stage constraints usually appear in a covariant form. One therefore expects manifest covariance of the extended formulation.}.

In the second part of this work, we discuss Lagrangian local symmetries.
The Dirac procedure, being applied to the extended Lagrangian $\tilde L$, stops at the second stage. That is $\tilde L$ represents the formulation with at most secondary constraints. It allows one to find a complete irreducible set of local symmetries of $\tilde L$, which turns out to be much easier task then those for $L$. The reason is as follows.
Appearance of some $N$-th stage first-class constraint in the Hamiltonian formulation for $L$ implies [3, 5, 7], that $L$ has the local symmetry of ${\stackrel{(N-1)}{\epsilon}}$-type
\begin{eqnarray}\label{01}
\delta q=\epsilon R^{(0)}+\dot\epsilon R^{(1)}+\ddot\epsilon R^{(2)}+\ldots+
{\stackrel{(N-1)}{\epsilon}}R^{(N-1)}.
\end{eqnarray}
Here ${\stackrel{(k)}{\epsilon}}$ $\equiv$ $\frac{d^k\epsilon}{d\tau^k}$, $\epsilon(\tau)$ is the local symmetry parameter and the set $R^{(k)}(q, \dot q, \ldots)$ represents gauge generator.
Replacing $L$ with $\tilde L$, one deal with the formulation with at most second-stage constraints. So, the symmetry (\ref{01}) of $L$ "decomposes" into $N$ simple $\dot{\epsilon}$-type symmetries of $\tilde L$
\begin{eqnarray}\label{02}
\delta q=\tilde\epsilon^I \tilde R_I^{(0)}+\dot{\tilde\epsilon}^I \tilde R_I^{(1)}, \qquad
I=1, 2, \ldots , N.
\end{eqnarray}
Technically, our procedure here is based on the works [4, 3]. In [4], it was observed that symmetries of
$H_{ext}$ can be written in a closed form. In [3], this observation was used to formulate a procedure for restoration of symmetries of the corresponding complete
Hamiltonian action\footnote{While the algorithm suggested is relatively simple, some of its points remain unclarified. In particular, the completeness and irreducibility of the symmetries of the complete Hamiltonian were not demonstrated so far [5]. The Lagrangian symmetries have not been discussed. We also point out that the analysis of a general case (when both first and second class constraints are present) turns out to be a much more complicated issue [5-9].}. For our case, the symmetries of $\tilde H$ can be immediately written down as it has the structure similar to that of $H_{ext}$. By a direct computation we then demonstrate that the Lagrangian counterparts of the symmetry transformations leave the extended Lagrangian invariant.

This work is organized as follows. With the aim to fix our notations, we describe in Section 2 the
Hamiltonization procedure for a Lagrangian theory with first class constraints up to $N$-th stage.
In Section 3, we introduce the improved extended Hamiltonian and obtain the underlying Lagrangian.
We then demonstrate the equivalence of the initial and the extended Lagrangian formulations.
In Section 4, we demonstrate one of the advantages of the extended formulation by finding its complete
irreducible set of local symmetries. The procedure is illustrated with an
example of a model with fourth-stage constraints.

\section{Initial formulation with higher stage constraints}
Let $L(q^A, \dot q^B)$ be the Lagrangian of a singular theory:
$rank\frac{\partial^2 L}{\partial\dot q^A\partial\dot q^B}=[i]<[A]$, defined on
configuration space $q^A, A=1, 2, \ldots , [A]$. From the beginning, it is convenient to rearrange the  initial
variables in such a way
that the rank minor is placed in the upper left corner. Then one has $q^A=(q^i, q^\alpha)$,
$i=1, 2, \ldots , [i]$, ~
$\alpha=1, 2, \ldots , [\alpha]=[A]-[i]$, where
$\det\frac{\partial^2 L}{\partial\dot q^i\partial\dot q^j}\ne 0$.

Let us construct the Hamiltonian formulation for the theory. To fix our notations, we carry out the
Hamiltonization procedure in some details. One introduces conjugate
momenta according to the equations $p_A=\frac{\partial L}{\partial\dot q^A}$, or
\begin{eqnarray}\label{1}
p_i=\frac{\partial L}{\partial\dot q^i},
\end{eqnarray}
\begin{eqnarray}\label{1.1}
p_\alpha=\frac{\partial L}{\partial\dot q^\alpha}.
\end{eqnarray}
They are considered as algebraic equations for determining velocities $\dot q^A$.
According to the rank condition, $[i]$ equations (\ref{1})
can be resolved with respect to $\dot q^i$. Let us denote the solution as
\begin{eqnarray}\label{2}
\dot q^i=v^i(q^A, p_j, \dot q^\alpha).
\end{eqnarray}
It can be substituted into remaining $[\alpha]$ equations for the momenta (\ref{1.1}). By construction, the
resulting expressions do not depend on $\dot q^A$ and are called primary constraints
$\Phi_\alpha(q, p)$ of the Hamiltonian formulation. One finds
\begin{eqnarray}\label{3}
\Phi_\alpha\equiv p_\alpha-f_\alpha(q^A, p_j)=0,
\end{eqnarray}
where
\begin{eqnarray}\label{4}
f_\alpha(q^A, p_j)\equiv\left.\frac{\partial L}{\partial\dot q^\alpha}
\right|_{\dot q^i=v^i(q^A, p_j, \dot q^\alpha)}.
\end{eqnarray}
The equations (\ref{1}) (\ref{1.1}) are thus equivalent to the system (\ref{2}), (\ref{3}).

Next step of the Hamiltonian procedure is to introduce an extended phase space parameterized by the
coordinates $q^A, p_A, v^\alpha$, and to define a complete Hamiltonian $H$ according to the rule
\begin{eqnarray}\label{5}
H(q^A, p_A, v^\alpha)=H_0(q^A, p_j)+v^\alpha\Phi_\alpha(q^A, p_j, p_\alpha),
\end{eqnarray}
where
\begin{eqnarray}\label{6}
H_0=\left.(p_i\dot q^i-L+ \dot q^\alpha\frac{\partial L}{\partial \dot q^\alpha})\right|
_{\dot q^i\rightarrow v^i(q^A, p_j, \dot q^\alpha)}.
\end{eqnarray}
Then the following system of equations on this space
\begin{eqnarray}\label{7}
\dot q^A=\{q^A, H\}, \qquad \dot p_A=\{p_A, H\}, \qquad
\Phi_\alpha(q^A, p_B)=0,
\end{eqnarray}
is equivalent to the Lagrangian equations following from $L$, see [2]. Here $\{ , \}$ denotes
the Poisson bracket. Let us point that equations for $\dot q^i$ of the system (\ref{7}) coincide,
modulo notations, with Eq. (\ref{2}), where $\dot q^\alpha$ are replaced by $v^\alpha$ (see [10] for more details).
This fact will be used in the next section.

It may happen that
the system (\ref{7}) contains in reality more then $[\alpha]$ algebraic equations. Actually, the derivative of the
primary constraints with respect to time implies the so called second-stage equations
as algebraic consequences of the system (\ref{7}): $\{\Phi_{\alpha}, H\}$ $=$ $0$.
Let us suppose that on-shell these expressions
do not involve the Lagrangian multipliers $v^\alpha$. Functionally independent equations of the system, if any,
represent then secondary Dirac constraints $\Phi^{(2)}_{\alpha_2}(q^A, p_j)=0$. They may imply third-stage constraints,
and so on. We suppose that the theory has
constraints up to $N$-th stage, $N\geq 2$. Higher stage constraints (that is those of second stage, third stage,
$\ldots$) are denoted collectively by $T_a(q^A, p_j)=0$. Then the complete system of constraints is
$G_I\equiv(\Phi_\alpha, , T_a)$. In this work we restrict ourselves to the case of a theory
with first class constraints only
\begin{eqnarray}\label{8}
\{G_I, G_J\}=c_{IJ}{}^{K}(q^A, p_j)G_K, \qquad \{G_I, H_0\}=b_{I}{}^J(q^A, p_j)G_J,
\end{eqnarray}
where $c, ~ b$ are phase space functions. Since the quantities on the l.h.s. of these equations are at most linear in
$p_\alpha$, one has: $c_{IJ}{}^\alpha=0$, ~ $b_{I}{}^\alpha=0$.

\section{Reconstruction of higher stage constraints into at most secondary ones}
As it was mentioned in the introduction, the extended Hamiltonian
\begin{eqnarray}\label{8.1}
H_{ext}(q^A, p_A, v^\alpha, \lambda^a)=H_0(q^A, p_j)+v^\alpha\Phi_\alpha(q^A, p_j, p_\alpha)+
\lambda^aT_a(q^A, p_j),
\end{eqnarray}
cannot be obtained as the complete Hamiltonian for some Lagrangian. Actually, $T_a$ can be chosen in a form resolved with respect to the momenta $p_a$: $T_a$ $=$ $p_a$ $-$ $t_a(q^A, p')$. Then it is clear that the Eq. (\ref{8.1}) does not have the desired form (\ref{5}), since $H_0$ from (\ref{8.1}) generally depends on $p_a$. To improve this, let us introduce the following quantity
\begin{eqnarray}\label{9}
\tilde H(q^A, \tilde p_A, s^a, \pi_a, v^\alpha, v^a)=\tilde H_0(q^A, \tilde p_j, s^a)+v^\alpha\Phi_\alpha(q^A, \tilde p_B)+v^a\pi_a,
\end{eqnarray}
defined on the space parameterized by the coordinates $q^A$, $\tilde p_A$, $s^a$, $\pi_a$,
$v^\alpha$, $v^a$.
In Eq. (\ref{9}) it was denoted
\begin{eqnarray}\label{10}
\tilde H_0=H_0(q^A, \tilde p_j)+s^aT_a(q^A, \tilde p_j).
\end{eqnarray}
The functions $\Phi_\alpha$, $H_0$, $T_a$ are taken from the initial formulation.
The improved Hamiltonian (\ref{9}) has the desired structure and can be obtained as the complete Hamiltonian of some Lagrangian $\tilde L(q^A, \dot q^A, s^a)$ defined on a configuration space with the coordinates $q^A$, $s^a$. As it will be shown below, the quantity $\tilde H_0$ turns out to be the corresponding Hamiltonian, while the equations $\Phi_\alpha$ $=$ $0$, ~ $\pi_a$ $=$ $0$ ($\pi_a$ are conjugate momenta for $s^a$) represent the primary constraints.
Due to the special form of the Hamiltonian
(\ref{10}), the preservation in time of the primary constraints
implies that all the higher stage constraints $T_a$ of the initial theory appear as secondary constraints for
the theory $\tilde L$. The Dirac procedure for $\tilde L$ stops at the second stage.
Besides, we demonstrate that the formulations $L$ and $\tilde L$ are equivalent.

Let us start from construction of the extended Lagrangian.
One writes the equations
\begin{eqnarray}\label{11}
\dot q^i=\frac{\partial\tilde H}{\partial\tilde p_i}=
\frac{\partial H_0}{\partial\tilde p_i}+s^a\frac{\partial T_a}{\partial\tilde p_i}-
v^\alpha\frac{\partial f_\alpha}{\partial\tilde p_i}.
\end{eqnarray}
They can be resolved algebraically with respect to $\tilde p_i$ in a neighborhood of the
point $s^a=0$. Actually, Eq. (\ref{11}) with $s^a=0$ is
$\dot q^i=\frac{\partial H}{\partial\tilde p_i}$, that is
the equation $\dot q^i=v^i(q^A, p_j, \dot q^\alpha)$ of the initial theory. Its solution exists and
is written in Eq. (\ref{1}). Hence
$\det\frac{\partial^2\tilde H}{\partial\tilde p_i\partial\tilde p_j}\ne 0$ at the
point $s^a=0$. Then the same is true in some vicinity of this point, and Eq. (\ref{11})
thus can be resolved. Let us denote the solution as
\begin{eqnarray}\label{12}
\tilde p_i=\omega_i(q^A, \dot q^i, v^\alpha, s^a).
\end{eqnarray}
By construction, one has the identities
\begin{eqnarray}\label{13}
\left.\omega_i\right|_{\dot q^i=\frac{\partial\tilde H}{\partial\tilde p_i}}\equiv\tilde p_i, \qquad
\left.\frac{\partial\tilde H}{\partial\tilde p_i}\right|_{\tilde p_i=\omega_i}\equiv\dot q^i,
\end{eqnarray}
as well as the following property of $\omega$
\begin{eqnarray}\label{14.0}
\left.\omega_i(q^A, \dot q^i, v^\alpha, s^a)\right|_{s^a=0, v^\alpha\rightarrow\dot q^\alpha}=
\frac{\partial L}{\partial\dot q^i}.
\end{eqnarray}
Below we use the notation
\begin{eqnarray}\label{14.3}
\left.\omega_i(q^A, \dot q^i, v^\alpha, s^a)\right|_{v^\alpha\rightarrow\dot q^\alpha}\equiv\omega_i(q, \dot q, s),
\end{eqnarray}
Now, on the configuration space parameterized by $q^A, s^a$, we define the extended
Lagrangian according to the rule
\begin{eqnarray}\label{14}
\tilde L(q^A, \dot q^A, s^a)= \qquad \qquad \qquad \qquad \qquad \cr
\left.\left(\omega_i\dot q^i+f_\alpha(q^A, \omega_j)\dot q^\alpha-
H_0(q^A, \omega_j)-s^aT_a(q^A, \omega_j)\right)\right|_{\omega(q, \dot q, s)}.
\end{eqnarray}
As compared with the initial Lagrangian, $\tilde L$ involves the new variables $s^a$, in a number
equal to the number of higher stage constraints $T_a$. Considering $\tilde L$ as a function
of $\omega$, one finds $\left.\frac{\partial\tilde L}{\partial\omega_i}
\right|_{\omega(q, \dot q, s)}$ $=$ $\left.\left(\dot q^i-\frac{\partial\tilde H}{\partial\tilde p_i}\right)
\right|_{\tilde p=\omega(q, \dot q, s)}$ $=$ $0$, according to the identity (\ref{13}). Thus the new
Lagrangian obeys the property
\begin{eqnarray}\label{14.1}
\left.\frac{\partial\tilde L}{\partial\omega_i}
\right|_{\omega(q, \dot q, s)}=0.
\end{eqnarray}
Note also that the Lagrangian counterparts $ T_a(q, \omega)$ of the higher-stage constraints enter into the extended Lagrangian in a manifest form. These properties will be crucial for the discussion of local symmetries in the next section.

Using Eq. (\ref{6}), $\tilde L$ can be written also in terms of the initial Lagrangian
\begin{eqnarray}\label{15}
\tilde L(q^A, \dot q^A, s^a)=L(q^A, v^i(q^A, \omega_j, \dot q^\alpha), \dot q^\alpha)+ \cr
\omega_i(\dot q^i-v^i(q^A, \omega_j, \dot q^\alpha))-s^aT_a(q^A, \omega_i), \quad
\end{eqnarray}
where the functions $v^i, \omega_i(q, \dot q, s)$ are given by Eqs. (\ref{2}), (\ref{14.3}).

Following the standard prescription [1, 2], let us construct the Hamiltonian formulation
for $\tilde L$. Using Eq. (\ref{14}), one finds conjugate momenta for $q^A, s^a$
\begin{eqnarray}\label{16}
\tilde p_i=\frac{\partial\tilde L}{\partial\dot q^i}=\omega_i(q^A, \dot q^A, s^a), \qquad
\tilde p_\alpha=\frac{\partial\tilde L}{\partial\dot q^\alpha}=f_\alpha(q^A, \omega_j), \cr
\pi_a=\frac{\partial\tilde L}{\partial\dot s^a}=0. \qquad \qquad \qquad \quad
\end{eqnarray}
Due to the identities (\ref{13}), these expressions can be rewritten in the equivalent form
\begin{eqnarray}\label{17}
\dot q^i=\frac{\partial\tilde H}{\partial\tilde p_i}, \qquad
\tilde p_\alpha=f_\alpha(q^A, \tilde p_j), \qquad \pi_a=0.
\end{eqnarray}
Thus the velocities $\dot q^i$ have been
determined, while as the primary constraints there appear $\pi_a=0$, and the primary constraints
$\Phi_\alpha=0$ of the initial theory. One finds the Hamiltonian $\tilde H_0$
\begin{eqnarray}\label{18}
\tilde H_0=\tilde p_A\dot q^A+\pi_a\dot s^a-\tilde L=H_0+s^aT_a(q^A, \tilde p_j),
\end{eqnarray}
so the complete Hamiltonian $\tilde H$ is given by Eq. (\ref{9}). Further, the preservation in time of the
primary constraints $\pi_a=0$ implies the equations $T_a=0$. Hence all the higher stage constraints of the initial
formulation appear now as secondary constraints. Preservation in time of the primary constraints $\Phi_\alpha$
leads to the equations $\{\Phi_\alpha, \tilde H\}=
\{\Phi_\alpha, H\}+s^ac_{\alpha a}{}^IG_I\approx\{\Phi_\alpha, H\}=0$, that is to the second stage
equations of the initial formulation. Hence, as before, they imply the secondary
constraints $\Phi^{(2)}_{\alpha_2}=0$; the latter appeared already as a part of the set $T_a=0$. The
Dirac procedure stops at the second stage. Owing to the structure
of the gauge algebra (\ref{8}), there are neither higher stage constraints nor equations for determining
the multipliers $v^\alpha, v^a$.

Let us compare the theories $\tilde L$ and $L$. The dynamics of the theory $\tilde L$ is governed by the
Hamiltonian equations
\begin{eqnarray}\label{19}
\dot q^A=\{q^A, H\}+s^a\{q^A, T_a\}, \qquad \dot{\tilde p}_A=\{\tilde p_A, H\}+s^a\{\tilde p_A, T_a\}, \cr
\dot s^a=v^a, \qquad \dot\pi_a=0, \qquad \qquad \qquad \qquad \quad
\end{eqnarray}
as well as by the constraints
\begin{eqnarray}\label{20}
\pi_a=0, \qquad \Phi_\alpha=0, \qquad T_a=0.
\end{eqnarray}
Here $H$ is the complete Hamiltonian of the initial theory (\ref{5}), and the Poisson bracket is defined
on the phase space $q^A, s^a, p_A, \pi_a$.
Let us make a partial fixation of the gauge by imposing the equations $s^a=0$ as gauge conditions for the
constraints $\pi_a=0$. Then $(s^a, \pi_a)$-sector of the theory disappears, while the remaining equations
in (\ref{19}), (\ref{20}) coincide with those of the initial theory\footnote{In a more rigorous treatment,
one writes Dirac brackets corresponding to the equations $\pi_a=0, s^a=0$. Then the latter can be used before
the computation of the brackets, that is the variables $s^a, \pi_a$ can be omitted. For the remaining phase-space
variables $q^A, p_A$, the Dirac bracket coincides with the Poisson one.} $L$. Let us remind that $\tilde L$
has been constructed in some vicinity of the point $s^a=0$. Admissibility of the gauge $s^a=0$ then guarantees
a self consistency of the construction. Thus $L$ represents one of the
gauges for $\tilde L$, which proves equivalence of the two formulations.

\section{Restoration of Lagrangian local symmetries from known Hamiltonian constraints}

Search for constructive and simple method of finding all the local symmetries of a given Lagrangian action is an interesting problem under investigation [5-9].
Here we demonstrate one of the advantages of the extended Lagrangian formulation developed in the previous section:
it is much easier to find local symmetries of $\tilde L$ than those of $L$.
The extended Lagrangian $\tilde L$ implies $[\alpha]+[a]$ primary first class constraints, so one expects the
same number of independent local symmetries in this formulation.
The symmetries can be easily found in an explicit form in terms of the constraints $G_I$ of the initial
formulation.

{\bf Symmetries of Hamiltonian action.} We start from discussion of local symmetries for the  Hamiltonian action, which corresponds to $\tilde L$
\begin{eqnarray}\label{21}
S_{\tilde H \tilde L}=\int d\tau(\tilde p_A\dot q^A+\pi_a\dot s^a-\tilde H)= \qquad \qquad \cr
\int d\tau(\tilde p_A\dot q^A+\pi_a\dot s^a-H_0(q^A, \tilde p_j)-s^aT_a(q^A, \tilde p_j)- \cr
v^\alpha\Phi_\alpha(q^A, \tilde p_B)-v^a\pi_a). \qquad \qquad \qquad
\end{eqnarray}
Let us consider variation of $S_{\tilde H\tilde L}$ under infinitesimal transformation
$\delta_I q^A=\epsilon^I\{q^A, G_I\}$, ~ $\delta_I\tilde p_A=\epsilon^I\{\tilde p_A, G_I\}$,
where $\epsilon^I$ are the parameters, and $I$ stands for any fixed $\alpha$ or $a$.
It implies (modulo to total derivative terms which we omit in subsequent computations)
$\delta(\tilde p_A\dot q^A)=\dot\epsilon^IG_I$, and $\delta A(q, \tilde p)=\epsilon^I\{A, G_I\}$ for any
function $A(q, \tilde p)$. Owing to these relations,
variation of $S_{\tilde H\tilde L}$ is proportional to $\Phi_\alpha, T_a$, so it can be canceled by
appropriate transformation of $v^\alpha, s^a$. In turn, the transformation of $s^a$ implies
$\delta(\pi_a\dot s^a)=\pi_a(\delta s^a)^{.}$, which can be canceled by variation of $v^a$:
$\delta v^a=(\delta s^a)^{.}$. Direct computations show, that the following
transformations\footnote{Transformation law for $v^\alpha$ turns out to be
$\delta v^\alpha=\dot\epsilon^\alpha+\epsilon^I b_I{}^\alpha-s^b\epsilon^I c_{b I}{}^\alpha-
v^\beta\epsilon^I c_{\beta I}{}^\alpha$, but the last three terms vanish, see end of Section 2.}:
\begin{eqnarray}\label{22}
\delta_I q^A=\epsilon^I\{q^A, G_I\}, \qquad \delta_I\tilde p_A=\epsilon^I\{\tilde p_A, G_I\}, \qquad \qquad \quad ~ ~ \cr
\delta_I s^a=\dot\epsilon^a\delta_{aI}+\epsilon^I b_I{}^a-s^b\epsilon^I c_{b I}{}^a-
v^\beta\epsilon^I c_{\beta I}{}^a, \qquad \delta_I\pi_a=0, \cr
\delta_I v^\alpha=\dot\epsilon^\alpha\delta_{\alpha I}, \qquad \delta_I v^a=(\delta_I s^a)^{.} \qquad \qquad \qquad \qquad \qquad ~ ~
\end{eqnarray}
leave invariant, modulo to a surface term, the Hamiltonian action (\ref{21}). Here $b, c$ are structure
functions of the gauge algebra (\ref{8}). Thus all the constraints $G_I$
of initial formulation turn out to be infinitesimal generators of the transformations in
$q^A, \tilde p_A$-subspace
of the phase space.

{\bf Symmetries of the extended Lagrangian action.} Let us demonstrate that it implies invariance
of the Lagrangian
action $S_{\tilde L}=\int d\tau\tilde L$ under the transformations
\begin{eqnarray}\label{23}
\delta_I q^A=\epsilon^I\left.\{q^A, G_I\}\right|_{\tilde p\rightarrow\omega(q, \dot q, s)}, ~\Leftrightarrow ~
\left\{
\begin{array}{ccc}
\delta_I q^\alpha & = & \epsilon^\alpha\delta_{\alpha I},\\
\delta_I q^i & = & \epsilon^I\left.\frac{\partial G_I}{\partial\tilde p_i}
\right|_{\tilde p\rightarrow\omega(q, \dot q, s)};\\
\end{array}
\right.
\cr
\delta_I s^a=\left.\left(\dot\epsilon^a\delta_{aI}+\epsilon^I b_I{}^a-s^b\epsilon^I c_{b I}{}^a-
\dot q^\beta\epsilon^I c_{\beta I}{}^a\right)\right|_{\tilde p\rightarrow\omega(q, \dot q, s)}, \qquad \qquad  ~ ~
\end{eqnarray}
First one notes that variation of $\tilde L$
of Eq. (\ref{14}) under arbitrary transformation $\delta q^A, \delta s^a$ can be presented in the form
\begin{eqnarray}\label{24}
\delta\tilde L=-\dot\omega_i\delta q^i-\dot f_\alpha\delta q^\alpha+
\dot q^\alpha\left.\frac{\partial f_\alpha}{\partial q^A}\right|_{\omega(q, \dot q, s)}\delta q^A- \cr
\left.\frac{\partial H_0}{\partial q^A}\right|_{\omega(q, \dot q, s)}\delta q^A-
\delta s^aT_a-s^a\left.\frac{\partial T_a}{\partial q^A}\right|_{\omega(q, \dot q, s)}\delta q^A.
\end{eqnarray}
We have omitted the term
$\left.\frac{\partial\tilde L}{\partial\omega_i}\right|_{\omega(q, \dot q, s)}\delta\omega_i$, the
latter is zero as a consequence of the identity (\ref{14.1}). To see that $\delta\tilde L$ is total
derivative, we add the expression
$-\epsilon^I\left.\frac{\partial\tilde L}{\partial\omega_i}\right|_{\omega(q, \dot q, s)}
\left.\frac{\partial G_I}{\partial q^i}\right|_{\tilde p\rightarrow\omega}\equiv 0$ to r.h.s. of Eq. (\ref{24}).
With $\delta q^A$ given by Eq. (\ref{23}), one obtains after some algebra
\begin{eqnarray}\label{25}
\delta\tilde L=\left.\left[-\dot T_a\epsilon^a+
\dot q^\alpha\epsilon^\beta\left(\frac{\partial f_\alpha}{\partial q^\beta}-\frac{\partial f_\beta}{\partial q^\alpha}-
\left(\frac{\partial f_\alpha}{\partial q^i}\frac{\partial f_\beta}{\partial\tilde p_i}-\alpha\leftrightarrow\beta\right)\right)-
\right.\right.\cr
\left.\left.\left(s^a\epsilon^\alpha-\dot q^\alpha\epsilon^a\right)\left(\frac{\partial T_a}{\partial q^\alpha}+
\left(\frac{\partial f_\alpha}{\partial q^i}\frac{\partial T_a}{\partial\tilde p_i}-f\leftrightarrow T\right)\right)-
\qquad \qquad \right.\right. \cr
\left.\left.\epsilon^\alpha\left(\frac{\partial H_0}{\partial q^\alpha}-
\left(\frac{\partial H_0}{\partial q^i}\frac{\partial f_\alpha}{\partial\tilde p_i}-H_0\leftrightarrow f\right)\right)-
\qquad \qquad \qquad \right.\right.\cr
\left.\left.\epsilon^a\left(\frac{\partial H_0}{\partial q^i}\frac{\partial T_a}{\partial\tilde p_i}-H_0\leftrightarrow T\right)-
s^a\epsilon^b\left(\frac{\partial T_a}{\partial q^i}\frac{\partial T_b}{\partial\tilde p_i}-a\leftrightarrow b\right)-
\delta s^aT_a\right]\right|_{\tilde p\rightarrow\omega} \cr
\equiv\left.\left[\dot\epsilon^aT_a-\dot q^\alpha\epsilon^\beta\{\Phi_\alpha, \Phi_\beta\}+
(s^a\epsilon^\alpha-\dot q^\alpha\epsilon^a)\{\Phi_\alpha, T_a\}- \qquad \qquad \right.\right. \cr
\left.\left.\epsilon^\alpha\{H_0, \Phi_\alpha\}-\epsilon^a\{H_0, T_a\}-
s^a\epsilon^b\{T_a, T_b\}-\delta s^aT_a\right]\right|_{\tilde p\rightarrow\omega}, \qquad \cr
=\left.\left[\left(\dot\epsilon^a-\dot q^\alpha\epsilon^Ic_{\alpha I}{}^a+\epsilon^Ib_I{}^a-s^b\epsilon^Ic_{b I}{}^a-
\delta s^a\right)T_a\right]\right|_{\tilde p\rightarrow\omega}. \qquad \qquad
\end{eqnarray}
Then the variation of $s^a$ given by Eq. (\ref{23}) implies $\delta\tilde L=div$, as it has been stated.

It is interesting to discuss the special case, when the initial Lagrangian has at most secondary first class constraints. Then the extended Lagrangian has the same structure.
Nevertheless, namely for $\tilde L$ the local symmetries can be found in the closed
form (\ref{23}).

{\bf Symmetries of the initial action.} Let us consider a combination of the symmetries (\ref{23}):
$\delta\equiv\sum_I\delta_I$,
which obeys: $\delta s^a=0$, for all $s^a$. The Lagrangian $\tilde L(q, s)$ will be invariant under this symmetry
for any fixed value of $s^a$, in particular, for $s^a=0$. But owing to Eqs. (\ref{15}),
(\ref{14.0}), (\ref{2}), the reconstructed Lagrangian coincides with the initial one for $s^a=0$:
$\tilde L(q, 0)=L(q)$. So the initial action will be invariant under any transformation
\begin{eqnarray}\label{25.1}
\delta q^A=\left.\sum_{I}\delta_I q^A\right|_{s=0},
\end{eqnarray}
which obeys to the system $\left.\delta s^a\right|_{s=0}$, that is
\begin{eqnarray}\label{25.2}
\dot\epsilon^a+\epsilon^Ib_I{}^a-v^\beta\epsilon^Ic_{\beta I}{}^a=0, \qquad a=1, 2, \ldots, [a].
\end{eqnarray}
One has $[a]$ equations for $[\alpha]+[a]$ variables $\epsilon^I$. Similarly to Ref. [3], the equations can be solved by
pure algebraic methods, which give some $[a]$ of $\epsilon$ in terms of the remaining $\epsilon$ and their derivatives
of order less than $N$. It allows one to find $[\alpha]$ local symmetries of $L$. As it was already mentioned, the problem
here is to prove the completeness and the irreducibility of the set.

{\bf Example.} As an illustration, we look for local symmetries of a theory with fourth-stage constraints presented
in initial formulation. Let us consider the Lagrangian
\begin{eqnarray}\label{26}
L=\frac12(\dot x)^2+\xi (x)^2,
\end{eqnarray}
where $x^\mu(\tau), \xi(\tau)$ are configuration space variables, $\mu=0, 1, \ldots, n$,
$(x)^2\equiv\eta_{\mu\nu}x^\mu x^\nu$, $\eta_{\mu\nu}=(-, +, \ldots , +)$. The theory is manifestly
invariant under global transformations of $SO(1, n-1)$-group. As it will be seen, our procedure
preserves the invariance.

Denoting the conjugate momenta for
$x^\mu,  \xi$ as $p_\mu, p_{\xi}$, one obtains the complete Hamiltonian
\begin{eqnarray}\label{27}
H_0=\frac12p^2-\xi (x)^2+v_{\xi}p_{\xi},
\end{eqnarray}
where $v_{\xi}$ is multiplier for the primary constraint $p_\xi=0$. The complete system of constraints turns out
to be
\begin{eqnarray}\label{28}
\Phi_1\equiv p_{\xi}=0, \quad T_2\equiv x^2=0, \quad
T_3\equiv xp=0, \quad T_4\equiv p^2=0.
\end{eqnarray}
For the case, the variable $\xi$ plays the role of $q^\alpha$, while $x^\mu$ play the role of $q^i$.
The constraints obey to the gauge algebra (\ref{8}), with non vanishing coefficient functions being
\begin{eqnarray}\label{29}
c_{2 3}{}^2=-c_{3 2}{}^2=2, \qquad c_{2 4}{}^3=-c_{4 2}{}^3=4, \qquad c_{3 4}{}^4=-c_{4 3}{}^4=2; \cr
b_1{}^2=1, \qquad b_2{}^3=2, \qquad b_3{}^4=1, \qquad b_3{}^3=2\xi, \qquad b_4{}^3=4\xi.
\end{eqnarray}
Equations (\ref{9}), (\ref{12}) acquire the form
\begin{eqnarray}\label{30}
\tilde H=\frac12(1+2s^4)\tilde p^2-\xi x^2 +s^2(x)^2+s^3(xp)+v_\xi p_\xi+v^a\pi_a,
\end{eqnarray}
\begin{eqnarray}\label{31}
\tilde p^\mu=\frac{\dot x^\mu-s^3x^\mu}{1+2s^4}.
\end{eqnarray}
Using these equations, one writes the reconstructed Lagrangian (\ref{14})
\begin{eqnarray}\label{32}
\tilde L=\frac{1}{2(1+2s^4)}(\dot x^\mu-s^3x^\mu)^2+(\xi-s^2)(x^\mu)^2.
\end{eqnarray}
It suggests the following redefinition of variables: $1+2s^3\equiv e, \xi-s^2\equiv \xi_1$,
then the previous expression can be written in the form
\begin{eqnarray}\label{33}
\tilde L(e, \xi_1)=\frac{1}{2e}(\dot x^\mu-s^3x^\mu)^2+\xi_1(x^\mu)^2.
\end{eqnarray}
Note that the reconstructed Lagrangians (\ref{32}), (\ref{33}) remain invariant under $SO(1, n-1)$ global transformations.

The Lagrangian (\ref{32}) implies four primary constraints $p_\xi=0, \pi_a=0$, and the secondary constraints $T_a$
from Eq. (\ref{28}). The corresponding complete Hamiltonian is given by Eq. (\ref{30}). It has four irreducible local
symmetries, the corresponding parameters are denoted as $\epsilon^1, \epsilon^2, \epsilon^3, \epsilon^4$.
By using of Eqs. (\ref{22}), (\ref{29}), manifest
form of the symmetries can be written immediately as follows (we have omitted
the variations $\delta_iv^a=(\delta_is^a)^.$)
\begin{eqnarray}\label{34}
\delta_1\xi=\epsilon^1, \quad \delta_1s^2=\epsilon^1, \quad \delta_1v_\xi=\dot\epsilon^1, \qquad \qquad \qquad
\qquad \qquad ~
\end{eqnarray}
\begin{eqnarray}\label{35}
\delta_2\tilde p^\mu=-2\epsilon^2x^\mu, \quad \delta_2s^2=\dot\epsilon^2+2\epsilon^2s^3, \quad
\delta_2s^3=2\epsilon^2(1+2s^4);
\end{eqnarray}
\begin{eqnarray}\label{36}
\delta_3x^\mu=\epsilon^3 x^\mu, \quad \delta_3\tilde p^\mu=-\epsilon^3\tilde p^\mu, \qquad \qquad \qquad \qquad \qquad \qquad ~ \cr
\delta_3s^2=2\epsilon^3(\xi-s^2), \quad
\delta_3s^3=\dot\epsilon^3, \quad \delta_3s^4=\epsilon^3(1+2s^4); \qquad ~
\end{eqnarray}
\begin{eqnarray}\label{37}
\delta_4x^\mu=2\epsilon^4 \tilde p^\mu, \quad \delta_4s^3=4\epsilon^4(\xi-s^2), \quad
\delta_4s^4=\dot\epsilon^4-2\epsilon^4s^3. \quad
\end{eqnarray}
The corresponding symmetries for $\tilde L$ are obtained according to Eq. (\ref{23})
\begin{eqnarray}\label{38}
\delta_1\xi=\epsilon^1, \qquad \delta_1s^2=\epsilon^1;\qquad \qquad \qquad \qquad \qquad \qquad \qquad \qquad \quad
\end{eqnarray}
\begin{eqnarray}\label{39}
\delta_2s^2=\dot\epsilon^2+2\epsilon^2s^3,  \qquad \delta_2s^3=2\epsilon^2(1+2s^4);
\qquad \qquad \qquad \qquad ~
\end{eqnarray}
\begin{eqnarray}\label{40}
\delta_3x^\mu=\epsilon^3 x^\mu, ~ \delta_3s^2=2\epsilon^3(\xi-s^2), ~
\delta_3s^3=\dot\epsilon^3, ~ \delta_3s^4=\epsilon^3(1+2s^4);
\end{eqnarray}
\begin{eqnarray}\label{41}
\delta_4x^\mu=2\epsilon^4 \frac{\dot x^\mu-s^3x^\mu}{1+2s^4}, \quad \delta_4s^3=4\epsilon^4(\xi-s^2), \quad
\delta_4s^4=\dot\epsilon^4-2\epsilon^4s^3.
\end{eqnarray}
From these expressions one can write also the symmetries for $L(e, \xi_1)$ of Eq. (\ref{33}).
The symmetry (\ref{38}) disappears, since
$L(e, \xi_1)$ is constructed from it's gauge invariant variables.
The remaining symmetries acquire the form
\begin{eqnarray}\label{39.1}
\delta_2\xi_1=-\dot\epsilon^2-2\epsilon^2s^3,  \qquad \delta_2s^3=2e\epsilon^2;
\qquad \qquad \qquad \qquad ~ \quad
\end{eqnarray}
\begin{eqnarray}\label{40.1}
\delta_3x^\mu=\epsilon^3 x^\mu, ~ \delta_3\xi_1=-2\epsilon^3\xi_1, ~
\delta_3s^3=\dot\epsilon^3, ~ \delta_3e=2\epsilon^3e; \qquad ~ ~
\end{eqnarray}
\begin{eqnarray}\label{41.1}
\delta_4x^\mu=\frac{2\epsilon^4}{e}(\dot x^\mu-s^3x^\mu), \quad \delta_4s^3=4\epsilon^4\xi_1, \quad
\delta_4s^4=\dot\epsilon^4-2\epsilon^4s^3.
\end{eqnarray}
$\delta_4$-symmetry can be replaced by the combination $\delta_\epsilon\equiv
\delta(\epsilon^4=\frac12\epsilon e)+\delta(\epsilon^3=\epsilon s^3)+\delta(\epsilon^2=-\epsilon\xi_1)$,
the latter has more simple form as compare with (\ref{41.1})
\begin{eqnarray}\label{41.2}
\delta_\epsilon x^\mu=\epsilon\dot x^\mu, \quad \delta_\epsilon\xi_1=(\epsilon\xi_1)^., \quad
\delta_\epsilon s^3=(\epsilon s^3)^., \quad
\delta_\epsilon e=(\epsilon e)^.,
\end{eqnarray}
and represents reparametrization invariance.
As an independent symmetries of $L(e, \xi_1)$, one can take either Eqs. (\ref{39.1})-(\ref{41.1}), or
Eqs. (\ref{39.1}), (\ref{40.1}), (\ref{41.2}).

Since the initial Lagrangian $L$ implies unique chain of four first class constraints, one expects one local
symmetry of ${\stackrel{(3)}{\epsilon}}$-type [4]. It can be found according to defining
equations (\ref{25.2}), for the case
\begin{eqnarray}\label{42}
\begin{array}{ccccc}
\epsilon^1 & +\dot\epsilon^2 & +2\epsilon^3\xi & {} & =0, \\
{} & 2\epsilon^2 & +\dot\epsilon^3 & +4\epsilon^4\xi & =0, \\
{} & {} & \epsilon^3 &  +\dot\epsilon^4 & =0.
\end{array}
\end{eqnarray}
It allows one to find $\epsilon^1, \epsilon^2, \epsilon^3$ in terms of $\epsilon^4\equiv\epsilon$:
$\epsilon^1=-\frac12{\stackrel{(3)}{\epsilon}}+4\dot\epsilon\xi+2\epsilon\dot\xi$,
$\epsilon^2=\frac12\ddot\epsilon-2\epsilon\xi$, $\epsilon^3=-\dot\epsilon$. Then Eq. (\ref{25.1}) gives the
local symmetry of the Lagrangian (\ref{26})
\begin{eqnarray}\label{43}
\delta x^\mu=-\dot\epsilon x^\mu+2\epsilon\dot x^\mu, \qquad
\delta\xi=-\frac12{\stackrel{(3)}{\epsilon}}+4\dot\epsilon\xi+2\epsilon\dot\xi.
\end{eqnarray}

In resume, in this work we have presented a relatively simple way for finding the complete irreducible set of
local symmetries in the Lagrangian theory with first-class constraints. Instead of looking for the symmetries of the initial Lagrangian $L$, one can construct an equivalent
Lagrangian $\tilde L$, given by Eq. (\ref{15}), with at most secondary first-class constraints.
Local symmetries of $\tilde L$ can be immediately written down according to Eq. (\ref{23}).
To conclude, we point out that in a recent work [10] it was demonstrated that the primary
constraints, while convenient, turn out to be not necessary for the Hamiltonization procedure. So, one can say that for the theory with first class constraints there exists a formulation with secondary first-class constraints.

\section{Acknowledgments}
Author would like to thank the Brazilian foundations CNPq (Conselho Nacional de Desenvolvimento
Científico  e Tecnológico - Brasil) and FAPERJ for financial support.

\end{document}